\documentclass[prl,epsfig,10pt,a4paper,twocolumn, showpacs]{revtex4}
\usepackage[english]{babel}
\usepackage{graphicx,amssymb}

\begin{document}

\title{Tailoring of phononic band structures in colloidal crystals}

\author{J. Baumgartl}

\author{M. Zvyagolskaya}

\author{C. Bechinger}

\affiliation{2. Physikalisches Institut, Universit\"a{}t Stuttgart, 70550 Stuttgart, Germany}

\begin{abstract}
We report an experimental study of the elastic properties of a
two-dimensional (2D) colloidal crystal subjected to light-induced
substrate potentials. In agreement with recent theoretical
predictions \cite{Gruenberg2007} the phonon band structure of such
systems can be tuned depending on the symmetry and depth of the
substrate potential. Calculations with binary crystals suggest that
phononic band engineering can be also performed by variations of the
pair potential and thus opens novel perspectives for the fabrication
of phononic crystals with band gaps tunable by external
fields.
\end{abstract}


\maketitle

Materials with periodic variations in their elastic properties
receive currently much interest as phononic crystals. Analogue to
light propagation in photonic crystals
\cite{Joannopoulos2000,Velikov2002}, the transmitted spectrum of
sound waves traveling through phononic crystals exhibit band gaps
whose frequency is determined by the length scale on which the
elastic properties are modulated. Fabrication of phononic crystals
is achieved by embedding regular arrays of elastic inclusions in an
appropriate matrix. Experiments with millimeter-sized inclusions
\cite{Vasseur2001,Liu2000} or sub-micron-sized holes immersed in
elastic host materials \cite{Gorishnyy2005,Thomas2006} indeed show
acoustic band gaps at ultrasonic, sonic and hypersonic frequencies.
Recent Brillouin spectroscopy measurements on crystals made of
sub-micron-sized colloidal particles immersed in a liquid matrix
demonstrated band gaps in the hypersonic regime with the possibility
to tune the band gap by exchange of the surrounding liquid
\cite{Cheng2006}.

In this letter, we experimentally investigate the phononic
properties of a two-dimensional (2D) crystal of colloidal particles
being subjected to a periodic substrate potential. Depending on the
substrate strength and particle interactions, the phononic band
structure and thus the position and width of phononic band gaps can
be largely tuned. Because this concept applies not only to
micron-sized colloids but also to much smaller particles, this
suggests tailoring the phononic properties of atoms or molecules
confined to extended optical lattices \cite{Morsch2006}. 

Experiments were performed with an aqueous suspension of highly
charged polystyrene spheres with diameter $\sigma=2.4\mu{}m$ and a
polydispersity below $4\%$. The particles interact via a screened
Coulomb potential $\Phi(r)\propto{}Z^{2}\exp(-\kappa{}r)/r$ with
$Z\approx10000$ the renormalized surface charge and
$\kappa^{-1}\approx300nm$ the screening length. Both values were
determined according to a procedure described in
\cite{Baumgartl2006}. As sample cell we used a cuvette made of fused
silica with $200\mu{}m$ spacing between top and bottom plate which
was connected to a standard closed deionization circuit to maintain
stable ionic conditions during the measurements \cite{Wei1998}.
After sedimentation, the particles form a 2D colloidal system close
to the bottom plate.

One- and two-dimensional substrate potentials were created by
superimposing two perpendicularly aligned one-dimensional periodic
interference patterns created with a $P=5W$ frequency-doubled
$\textrm{Nd:YVO}_{4}$ laser ($\textrm{wave length}=532nm$). The
polarizations of the interference patterns were adjusted
perpendicular, therefore they act as two independent 1D periodic
substrate potentials for the colloidal particles
$U_{i}(x)=U_{0}^{i}\cos(2\pi{}x/d_{i})$ with $U_{0}^{i}$ the
potential amplitude, $d_{i}$ the lattice constant and $i=1,2$
\cite{Chowdhury1985}. Because the potential amplitude $U_{0}^{i}$
scales linearly with the intensities of the laser beams
\cite{Bechinger2001} this allows to continuously adjust the strength
of the underlying laser potentials. An additional laser beam was
scanned around the central region of the sample to create a boundary
box whose size could be continuously adjusted by a pair of
computer-controlled galvanostatically-driven mirrors. This allowed
us to adjust the particle density $\rho$ with an accuracy of
$\Delta\rho/\rho\approx0.01$ \cite{Brunner2002}.

\begin{figure}[b]
\includegraphics[width=7.5cm]{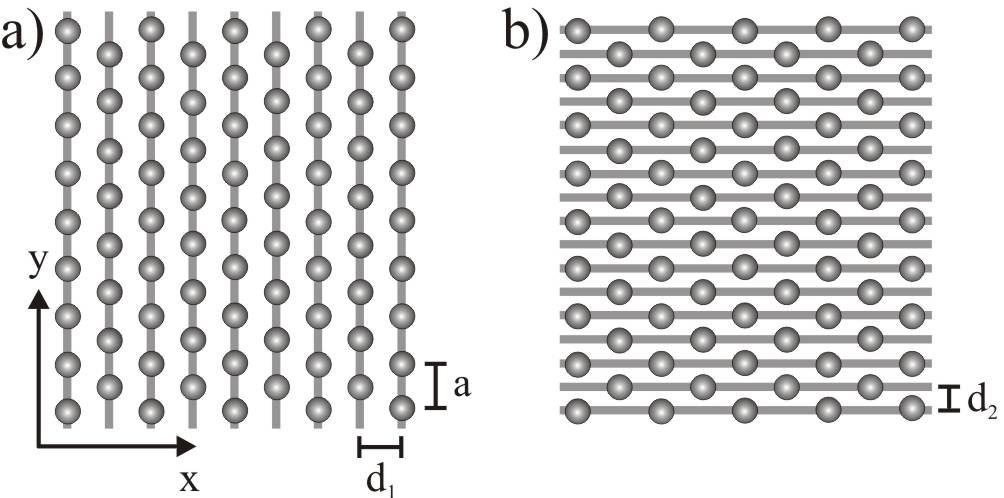}
\centering \caption{Illustration of the colloidal system subjected
to commensurate optical interference patterns. Gray dots represent
colloidal particles and gray lines indicate the minima of the
light-induced 1D periodic potentials. The lattice constants $d_{1}$
and $d_{2}$ of the 1D periodic potentials are chosen to be
commensurate with the mean particle distance $a$.} \label{Fig1}
\end{figure}

First, we continuously decreased the mean particle distance
$a=\sqrt{2/\sqrt{3}\rho}$ to approximately $4\mu{}m$ where under our
salt conditions spontaneous crystallization of the colloidal
monolayer occurs. Next, the lattice constants $d_{1}$ and $d_{2}$
were chosen to meet commensurate conditions, i.e.
$d_{1}=(\sqrt{3}/2)a\approx3.5\mu{}m$ and $d_{2}=a/2\approx2\mu{}m$
(see Fig. \ref{Fig1}). Particle positions were determined for
different combinations of interference pattern intensities
$\{U_{0}^{1},U_{0}^{2}\}$ from sequences consisting of several
thousand images using digital video microscopy at an acquisition
rate of $2$ frames per second \cite{Baumgartl2006}. From these data
we finally obtained the particle trajectories
$\vec{r}_{\mu}(t)=(x_{\mu}(t),y_{\mu}(t))$ with $\mu=1\dots{}N$ and
$t$ the time. To avoid boundary effects, we only considered the
central region of the field of view ($300\mu{}m\times200\mu{}m$)
containing $N\approx1000$ particles.

For the experimental determination of the phononic band structure we
followed the approach already introduced in
\cite{Gruenberg2007,Keim2004,Chaikin1995} which is based on the
analysis of the dynamical matrix. Briefly, the branch
$\lambda_{s}(\vec{q})$ of the phononic band structure with $\vec{q}$
the wave vector and polarization $s$ is given by the eigenvalue of
the dynamical matrix denoted as $D_{\alpha\beta}(\vec{q})$
($\alpha,\beta=x,y$). The latter is obtained from the measured
particle displacements $\vec{u}(\vec{R},t)$ relative to their
lattice sites $\vec{R}$
\begin{equation}
\label{eq:eqn1} D_{\alpha\beta}(\vec{q}) = k_{B}T/\langle{}u_{\alpha}^{\ast}(\vec{q},t)u_{\beta}(\vec{q},t)\rangle_{t}
\end{equation}
with $\vec{u}(\vec{q},t)$ the Fourier transform of $\vec{u}(\vec{R},t)$ and $\langle\dots\rangle_{t}$ the temporal average. Using the
equipartition theorem, this expression has been derived within the harmonic approximation of the potential energy
\begin{equation}
\label{eq:eqn2}H=1/2\sum_{\vec{q},\alpha,\beta}u_{\alpha}^{\ast}(\vec{q})D_{\alpha\beta}(\vec{q})u_{\beta}(\vec{q})
\end{equation}
which has been experimentally confirmed to be valid for colloidal systems \cite {Keim2004}. Note, that due to the absence of true long-range
order in 2D systems, the lattice sites $\vec{R}$ have to be determined through temporal averaging \cite{Mermin1968, remark2}. Within the
harmonic approximation the pair interaction between particles can be modeled as springs with the spring constant $k_{0}$ given by the second
derivative of the pair potential at the mean particle distance $k_{0}=[d^{2}\Phi(r)/dr^{2}]_{r=a}$. With the corresponding values $Z$ and
$\kappa$ taken from above we obtain $k_{0}\approx150k_{B}T/\sigma^{2}\approx1\cdot10^{-7}J/m^{2}$. The periodic substrates $U_{0}^{i}$ introduce
an additional set of springs which pin the particles to their lattice sites $\vec{R}$ with spring constants $k_{i}=U_{0}^{i}(2\pi/d_{1})^{2}$.
With the laser intensities used in our experiments we achieved maximum values of $k_{i}\approx2.5k_{0}$. Since the phononic band structure
depends only on the ratio between pair and particle-substrate interaction, in the following it will be expressed in units of $k_{0}$. We
confirmed the harmonic approximation to be valid over the entire range of observed displacements ($|u_{\alpha}|\leq1\mu{}m$) by determining the
effective single-particle potential $U_{eff}(u_{\alpha})/k_{B}T = -\ln[P(u_{\alpha})]$ with $P(u_{\alpha})$ the normalized particle-displacement
distribution.

\begin{figure}[b]
\includegraphics[width=7.5cm]{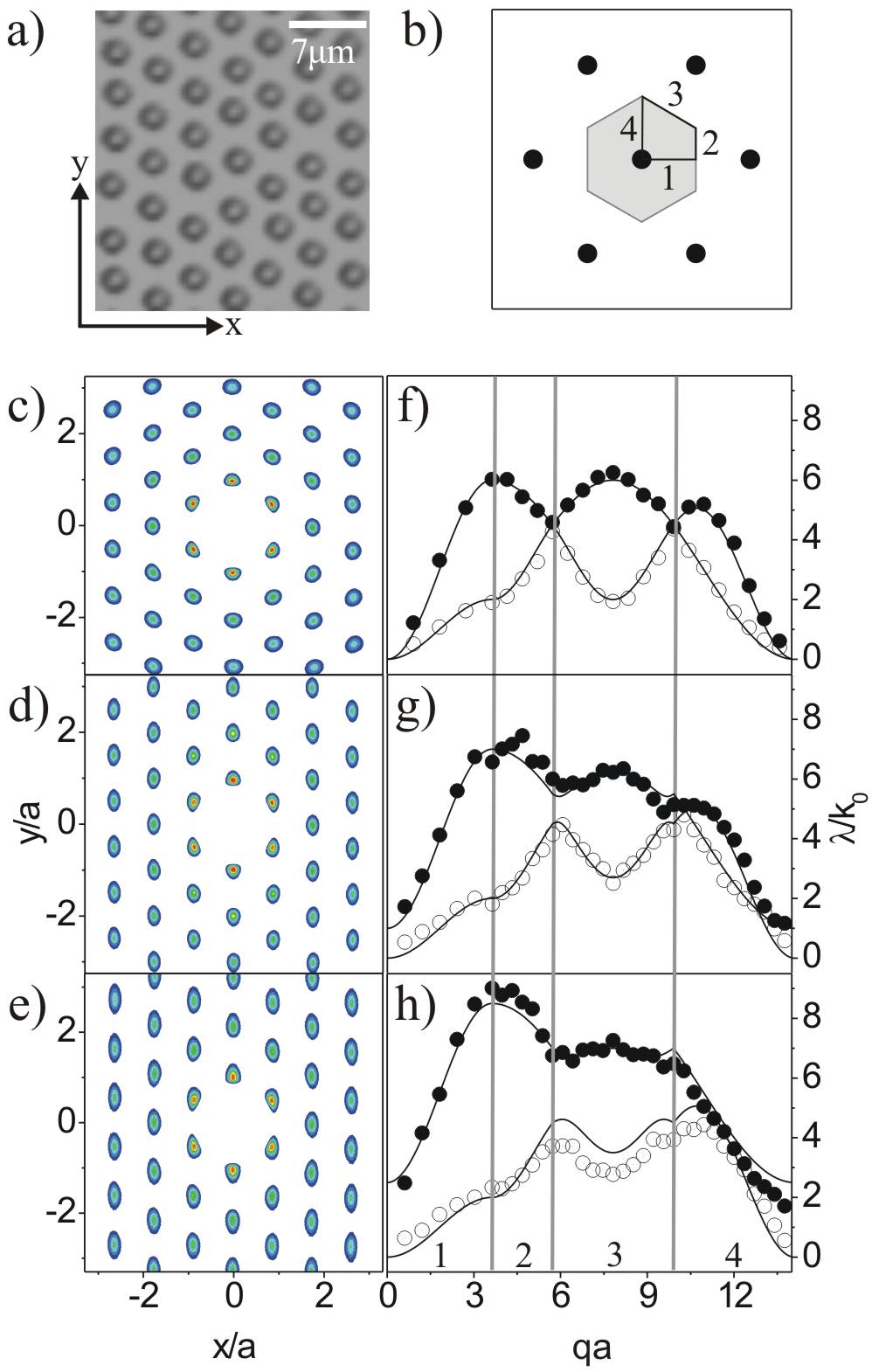}
\centering \caption{(color online) Experimental realization of a 2D
colloidal crystal on a 1D periodic substrate [see Fig.
\ref{Fig1}(a)]. a) Micrograph of the 2D colloidal crystal in the
central region of the field of view ($300\mu{}m\times200\mu{}m$). b)
Zero and first order Bragg peaks (black dots) and first Brillouin
zone (gray area). The black solid line represents the irreducible
path along which the phonon band structure is plotted and the path
sections are marked by numbers. c)-e) 2D correlation function
$g(x,y)$ and f)-h) phonon band structure $\lambda_{s}(q)$ along the
irreducible path for c),f) the spontaneous crystal, i.e.
$k_{1}/k_{0}=0$, d),g) $k_{1}/k_{0}=1$, and e),h) $k_{1}/k_{0}=2.5$.
Closed and open symbols refer to the upper and lower band,
respectively. Solid lines represent band structures calculated in
the framework of harmonic lattice dynamics \cite{Gruenberg2007}.}
\label{Fig2}
\end{figure}

\begin{figure}[b]
\includegraphics[width=7.5cm]{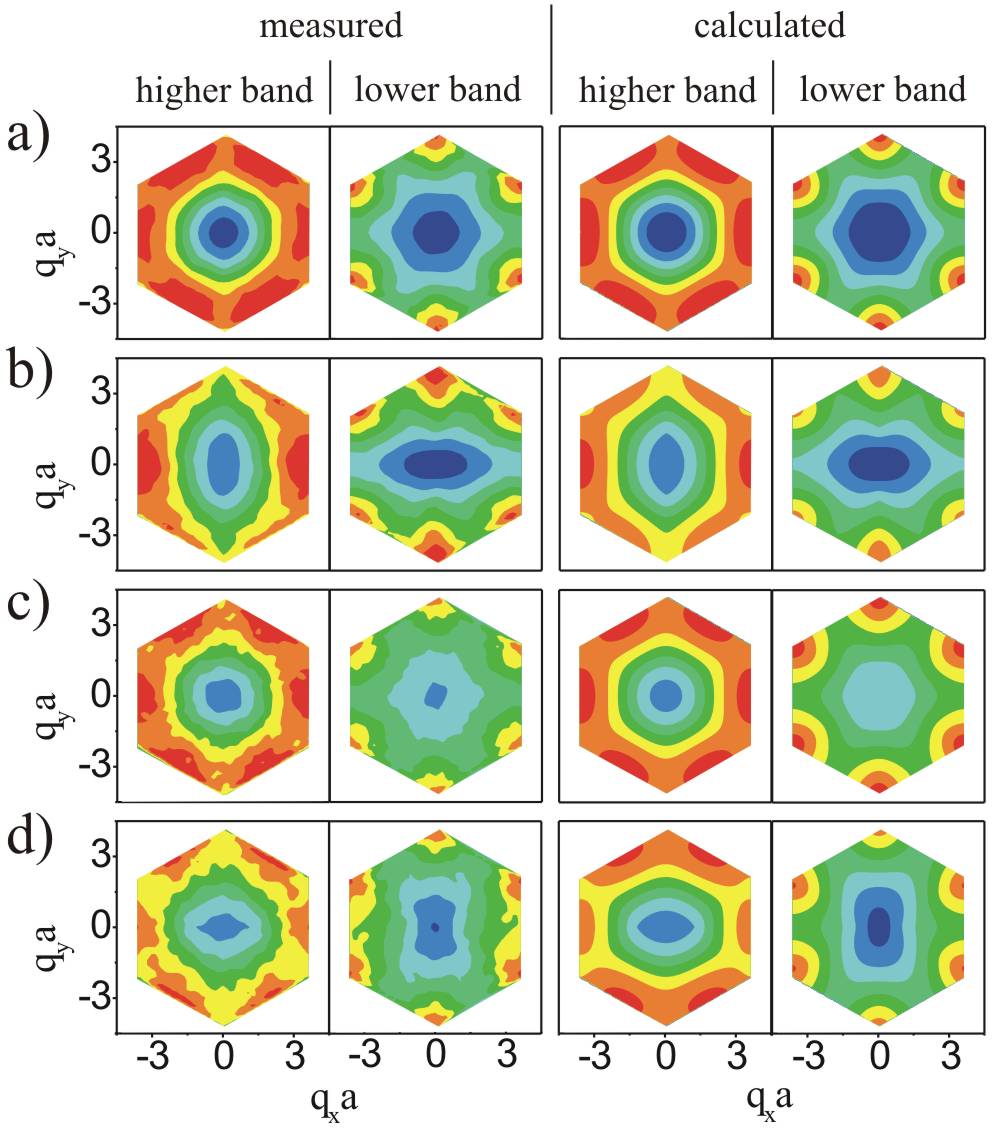}
\centering \caption{(color online) Experimental realization of a 2D
colloidal crystal in the presence of two 1D periodic substrates
rotated by $90^{\circ}$ (see Fig. \ref{Fig1}). The graphs are 2D
contour plots of the phonon band structure in the first Brillouin
zone derived from experimental (left column) and analytical data
(right column). a) $\{k_{1}/k_{0},k_{2}/k_{0}\}=\{0,0\}$. b)
$\{k_{1}/k_{0},k_{2}/k_{0}\}=\{1.4,0.2\}$. c)
$\{k_{1}/k_{0},k_{2}/k_{0}\}=\{1.5,1.5\}$. d)
$\{k_{1}/k_{0},k_{2}/k_{0}\}=\{0.5,1.5\}$. Experimental contour
plots were determined for approximately $800$ $\vec{q}$-values.
Calculated plots were obtained in the framwork of harmonic lattice
dynamics using approximately $3000$ $\vec{q}$-values
\cite{Gruenberg2007}.} \label{Fig3}
\end{figure}

\begin{figure}[b]
\includegraphics[width=7.5cm]{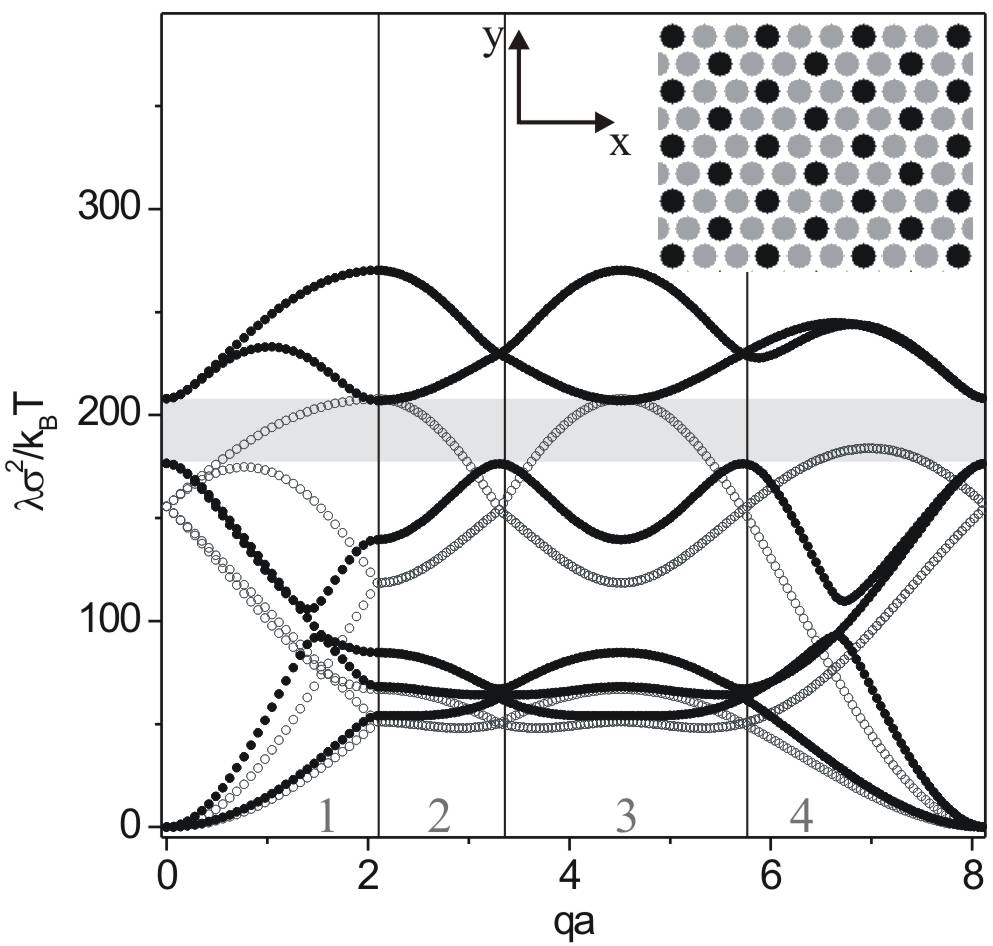}
\centering \caption{Phonon band structure (solid symbols) of a
paramagnetic 2D binary crystal (see inset) calculated in the
harmonic approximation. The black particles in the inset have a
magnetic susceptibility $20$ times larger than the susceptibility of
the gray particles. The magnetic field $\vec{B}$ is perpendicular to
the system plane. The band gap can be closed for $\vec{B}=\vec{0}$
and $\lambda>0$ if all the particles additionally interact via a
Lennard-Jones potential (open symbols).} \label{Fig4}
\end{figure}

We will start our discussion by first describing the results
obtained when only a single 1D substrate potential was applied
($U_{0}^{2}=0$) (see Fig. \ref{Fig2}). A typical snapshot of the
colloidal crystal is shown in Fig. \ref{Fig2}(a) (the 1D laser
potential which is aligned vertically is invisible because it is
blocked with optical filters). Fig. \ref{Fig2}(b) illustrates the
irreducible path along which the phonon band structure has been
analyzed. It corresponds to that route which circumvents that part
of the first Brillouin zone (gray area) being the smallest repeat
unit of the band structure. Due to the two-fold symmetry of the 1D
periodic substrate potential, this unit comprises a quarter of the
first Brillouin zone. Figures \ref{Fig2}(c)-(e) and (f)-(h) show the
2D pair correlation function $g(x,y)$ and the corresponding phonon
band structure for increasing laser intensities $k_{1}/k_{0}=0$
(spontaneous crystal) (c,f), $k_{1}/k_{0}=1$ (d,g) and
$k_{1}/k_{0}=2.5$ (e,h), respectively. Since the particles' motion
becomes more confined in the direction perpendicular to the
potential troughs, the fluctuations in x- and y-direction become
asymmetric with increasing $(k_{1}/k_{0})$ as this is observed in
the corresponding $g(x,y)$-plots.

From the above it is clear that changes in $g(x,y)$ also affect the
dynamical matrix and thus the phononic band structure. As we will
demonstrate in the following, those changes strongly depend on the
polarization of the phonons. Usually, the polarization $s$ refers to
longitudinal and transversal modes in the direction of high crystal
symmetry. Because we also determined the band structure in
directions with lower symmetry, in the following we will refer to
the different polarizations by their vertical position in Figs.
\ref{Fig2}(f)-(h) as lower ($s=l$) and upper ($s=u$) bands instead.
Interestingly, both bands are affected by the substrate potential in
a rather different way. We observe a pure shift of the upper band
$\lambda_{u}(\vec{q})$ (closed symbols) along the sections $1$ and
$2$ while the lower band $\lambda_{l}(\vec{q})$ (open symbols) is
not affected. This can be easily understood because
$\lambda_{u}(\vec{q})$ and $\lambda_{l}(\vec{q})$ have polarization
vectors in the $x$- and the $y$-direction along these sections,
respectively. As a consequence, $\lambda_{u}(\vec{q})$ is increased
by the spring constant $k_{1}/k_{0}$ of the periodic potential. The
bands have the same polarization along section $4$ of the
irreducible path. Therefore, the upper band is not influenced by the
periodic potential and the lower band is shifted by $k_{1}/k_{0}$.
We observe a different behavior along section $3$; here, both the
upper and the lower band are influenced because the respective
polarization does not point in the $x$- and the $y$-direction. The
upper band becomes largely deformed and is almost entirely flattened
in Fig. \ref{Fig2}(h). Because the group velocity
$\vec{v}(\vec{q})\propto\nabla\sqrt{\lambda(\vec{q})}$ vanishes in
those regions, this opens interesting perspectives regarding the
tailoring of acoustic and thermal properties of 2D crystals. We also
compared the experimentally determined band structures with
calculations based on the harmonic approximation
\cite{Gruenberg2007} using $k_{1}/k_{0}$ as a fit parameter, the
latter being in excellent agreement with the corresponding values
independently obtained from $\kappa^{-1}$ and $U_{0}^{1}$. The
calculated band structures are plotted as solid lines in Figs.
\ref{Fig2}(f)-(h).

Similar experiments have been also performed for 2D substrate
potentials ($\{U_{0}^{1},U_{0}^{2}\} \neq \{0,0\}$) and demonstrate
that both the upper and the lower band can be individually shifted
and deformed by the presence of the substrate potentials. Rather
than showing a specific path through the Brillouin zone, here we
have chosen a 2D representation of the phononic band structure which
is shown in Fig. \ref{Fig3} for different combinations of substrate
strengths. For comparison, we also show calculated 2D phonon band
structures obtained within the harmonic approximation using
$k_{1}/k_{0}$ and $k_{2}/k_{0}$ as fit parameters
\cite{Gruenberg2007}. For the substrate-free case where
$\{k_{1}/k_{0},k_{2}/k_{0}\}=\{0,0\}$, as expected we observe the
six-fold symmetry of a spontaneous crystal [Fig. \ref{Fig3}(a)]. In
Fig. \ref{Fig3}(b), the spring constants are
$\{k_{1}/k_{0},k_{2}/k_{0}\}=\{1.4,0.2\}$. Here, the symmetry breaks
down to a two-fold symmetry due to the dominating $k_{1}$. Fig.
\ref{Fig3}(c) corresponds to the situation where both interference
patterns have identical strengths
$\{k_{1}/k_{0},k_{2}/k_{0}\}=\{1.5,1.5\}$ and as a consequence the
six-fold symmetry is restored. Finally, Fig. \ref{Fig3}(d) shows the
case where $\{k_{1}/k_{0},k_{2}/k_{0}\}=\{0.5,1.5\}$. Similar as in
Fig. \ref{Fig3}(b) we find a two-fold symmetry but here rotated by
$90^{\circ}$ because of $k_{2}/k_{0}>k_{1}/k_{0}$.

After having discussed the possiblity to tailor phononic band
structures by subjecting monolayers to substrate potentials,
finally, we want to discuss whether phononic band engineering can
also be achieved by adjustable anisotropic pair potentials. Here, we
exemplarily consider a system with magnetic dipole-dipole
interactions
$\Phi_{m}(\vec{r})=\vec{d}^{2}/|\vec{r}|^{3}-3(\vec{d}\cdot\vec{r})^{2}/|\vec{r}|^{5}$.
For paramagnetic particles the dipole moment $\vec{d}$ scales as
$\vec{d}=\chi\vec{B}$ with $\chi$ the magnetic susceptibility and
$\vec{B}$ the magnetic field. The pair interaction becomes
anisotropic when the magnetic field is not applied perpendicular to
the plane of the 2D crystal. Our calculations based on the approach
in \cite{Gruenberg2007} indeed show that tuning of phononic bands is
possible under such conditions, however, we also find that
anisotropic pair potentials are not sufficient to create band gaps
in monodisperse systems. This is because only acoustic branches
exist in such systems where $\lambda(\vec{q})\to0$ for
$\vec{q}\to\vec{0}$. To overcome this limitation, one can create
additional optical branches ($\lambda(\vec{q})>0$ for
$\vec{q}=\vec{0}$). This is achieved by extending the unit cell of
the crystal by adding particles with different magnetic properties
to the system (black and gray spheres in the inset of Fig.
\ref{Fig4}). For our calculations we assumed a magnetic
susceptibility ratio $\chi_{1}/\chi_{2}=20$ between the black and
gray particles and the magnetic field $\vec{B}$ orientated
perpendicular to the system. The calculated phonon band structure is
shown in Fig. \ref{Fig4} as closed symbols and indeed shows four
additional optical branches. Most importantly, we observe a full
band gap whose frequency $\omega\propto\sqrt{\lambda}$ and width
scale linearly with the magnetic field. As a consequence, the band
gap can only be closed in the limit $\vec{B}\to\vec{0}$ where both
the pair interaction and the corresponding spring constants
$\lambda(\vec{q})$ vanish; therefore, we considered an additional
pair interaction which was exemplarily assumed as a Lennard-Jones
potential $\phi_{LJ}(r)\propto[(\sigma/r)^{12}-2(\sigma/r)^{6}]$.
Similar as above the band gap can be completely closed at
$\vec{B}=\vec{0}$ but now at finite frequencies, here determined by
the Lennard-Jones interaction (see open symbols in Fig. \ref{Fig4}).
This would allow to extend the possibility to tailor phononic band
structure also to situations where surface potentials can not be
modified. Preliminary calculations indicate that a similar behavior
can be observed in three-dimensional systems.

In conclusion, we have experimentally demonstrated that the phonon band structure of a 2D colloidal crystal can be greatly tuned by the strength
of periodic substrate potentials. Depending on the symmetry of the applied substrate potential different phonon polarizations can be tailored
rather independently. Calculations with binary crystals of paramagnetic particles with adjustable pair interactions indicate that phononic band
engineering can be also performed in situations where substrate potentials can not be tuned and thus opens novel perspectives for the
fabrication of phononic crystals with band gaps adjustable by external fields.

It is a great pleasure to acknowledge H.H. von Gr\"{u}nberg for
helpful ideas and stimulating discussions.

\noindent Electronic address: j.baumgartl@physik.uni-stuttgart.de

\bibliographystyle{apsrev}

\begin{thebibliography}{18}
\bibitem[{\citenamefont{remark2}(2007)}]{remark2}
\bibnamefont{A sliding time window $\Delta{}T$ was chosen and the lattice sites $\vec{R}(t)$ at time $t$ then followed as the temporal average
$\vec{R}(t)=\Delta{}T^{-1}\int_{t-\Delta{}T/2}^{t+\Delta{}T/2}dt'\vec{r}_{i}(t')$. The displacement $\vec{u}(\vec{R},t)$ from the lattice site
$\vec{R}$ is obtained straightforwardly as $\vec{u}(\vec{R},t) = \vec{r}_{i}(t)-\vec{R}(t)$. We performed the evaluation for the three values
$\Delta{}T=25s$, $40s$, and $60s$ and do not observe a dependence of our results on $\Delta{}T$ for the two larger values; therefore,
$\Delta{}T=40s$ is sufficiently long for accurate determination of the lattice sites $\vec{R}$.}
\end{thebibliography}

\end{document}